\documentclass[fleqn,usenatbib]{mnras}

\usepackage{newtxtext,newtxmath}

\usepackage[T1]{fontenc}
\usepackage{ae,aecompl}

\usepackage{graphicx}	
\usepackage{amsmath}	
\usepackage{amssymb}	
\usepackage{lipsum}
\usepackage{pdflscape}
\usepackage{afterpage}

\newcommand{\nustar}{\textit{NuSTAR}}

\newcommand{\xmm}{{\it XMM-Newton}}

\title[Modelling RMS Spectra I]{Modelling X-ray RMS spectra I: intrinsically variable AGN}

\author[M. L. Parker et al.]{M. L. Parker,$^{1}$\thanks{E-mail: mparker@sciops.esa.int}
W. N. Alston,$^2$
Z. Igo$^1$
and A. C. Fabian$^2$
\\
$^{1}$European Space Agency (ESA), European Space Astronomy Centre (ESAC), E-28691 Villanueva de la Ca\~{n}ada, Madrid, Spain\\
$^{2}$Institute of Astronomy, University of Cambridge, Madingley Road, Cambridge, CB3 0HA, UK\\
}

\date{Accepted XXX. Received YYY; in original form ZZZ}

\pubyear{2019}

\begin{document}
\label{firstpage}
\pagerange{\pageref{firstpage}--\pageref{lastpage}}
\maketitle

\begin{abstract}
We present simple \textsc{xspec} models for fitting excess variance spectra of AGN. Using a simple Monte-Carlo approach, we simulate a range of spectra corresponding to physical parameters varying, then calculate the resulting variance spectra. Starting from a variable power-law, we build up a set of models corresponding to the different physical processes that can affect the final excess variance spectrum.
We show that the complex excess variance spectrum of IRAS~13224-3809 can be well described by such an intrinsic variability model, where the power-law variability is damped by relativistic reflection and enhanced by an ultra fast outflow. The reflection flux is correlated with that of the power-law, but not perfectly. We argue that this correlation is stronger at high frequencies, where reverberation lags are detected, while excess variance spectra are typically dominated by low frequency variability.
\end{abstract}




\section{Introduction}

Emission from active galactic nuclei (AGN) is extremely variable, particularly in the X-ray band \citep[e.g.][]{Alston19}. This underlying variability process shows a characteristic log-normal distribution of source flux and a linear relation between the mean source flux and the RMS variability \citep{Uttley01}. This linear \emph{rms-flux} relation is observed over a broad range of 
timescales and requires that the underlying physical processes driving the variability must be multiplicative \citep{Uttley05}. These observed properties are most simply explained by the propagating fluctuations model \citep[e.g.][]{Lyubarskii97, King04, Arevalo06, Hogg16}, where variations on different timescales, with different amplitudes, from different annuli in the accretion disk are multiplied together. Any model seeking to explain AGN variability, and that of accreting compact objects in general, must explain these fundamental properties. 

Analysis of variability is not confined to calculating a single broad-band value. Excess variance spectra (often referred to as $F_\mathrm{var}$ or RMS spectra) are a common tool for quantifying spectral variability \citep[e.g.][]{Edelson02,Vaughan03_variability}. The principle is trivial: find the variance in a given energy band, subtract the expected variance for a constant lightcurve with the same count rate, and repeat with multiple energy bins until a complete spectrum is established. This can be done either in the time-domain, calculating the variance of lightcurves in specific energy bands, or in Fourier space by integrating the power spectrum for a given energy band between two frequencies. These are formally equivalent: the lowest frequency probed for a given lightcurve is simply the inverse of the lightcurve length, and the highest frequency is 1/(2dt), where dt is the timestep.

This technique has been used extensively with X-ray observations of AGN. \citet{Vaughan04} used it to show that the neutral Fe~K line and soft excess in MCG--06-30-15 are less variable than the continuum. \citet{Ponti06} calculated RMS spectra for NGC~4051 at different flux levels, demonstrating that the shape of the spectrum changes based on the contribution of constant emission components from distant reflection and photoionised gas. \citet{Middleton09} and \citet{Jin13} calculated RMS spectra in different frequency ranges for RE~J1034+396 and PG~1244+026, showing that the spectra are strongly frequency dependent.
\citet{Brenneman14} calculated a joint \xmm/\nustar\ variance spectrum for IC~4329A, showing that the variability damping from distant reflection extends up to the Compton hump.
\citet{Matzeu16} showed that the RMS spectrum of PDS~456 could be explained by a combination of intrinsic and absorption variability, with a constant distant reflection component. Recently, \citet{Parker17_irasvariability,Parker18_pds456} showed that the response of ionized absorption lines to the X-ray continuum results in a series of peaks in the RMS spectrum, which can be used to detect outflows.

Despite its prevelance, we argue that this tool has been under-used in X-ray astronomy as a quantitative measure. Analysis of RMS spectra is generally restricted to calculating such a spectrum, and sometimes comparing it to a simulated equivalent \citep[e.g.][]{Vaughan04} or constructing a custom model based on the best-fit to the count spectrum \citep[e.g.][]{Mallick17}. However, there is no reason in principle why RMS spectra cannot be fit with generally applicable models in an analogous way to flux or count spectra, and we believe the only reason this is not common practice is the lack of such usable models.

The aim of this paper is to remedy this lack, providing a set of basic models for fitting RMS spectra, accounting for various physical effects, along with suggestions and guidelines for their use. These models are publicly available, along with the code used to generate them, and we encourage other authors to try using them to model their RMS spectra in parallel to count spectra.


\section{Models}
\label{sec:models}
We take a Monte-Carlo approach to generating models. While some simple models could certainly be calculated analytically, this quickly becomes very difficult with more complex models and for our purposes it is simpler to use the same method for all models. We show a comparison between Monte-Carlo and analytic approaches in Appendix~\ref{sec:analytic}.

In brief, for each model we calculate a grid of spectra. For each grid point, we simulate 1000 model spectra using \textsc{xspec} 12.10.1f \citep{Arnaud96}, based on a combination of fixed parameters and parameters with a specified variance, then calculate the variance spectrum from these 1000 spectra. 

In general, we assume that one key parameter varies, and that some other parameters may be correlated it. These simple assumptions allow us to generate models that provide a remarkably good description of the observed excess variance spectra, as we show in section~\ref{sec:results}.

The exact parameters of each model table are given in Appendix~\ref{sec:tables}. The models themselves can be downloaded here: \url{http://bit.ly/fvar_models}.

\subsection{Power-law}
The simplest case that we consider is a variable power-law. Because RMS spectra are fractional, the absolute photon index and normalization are unimportant. The only parameters that are required to describe such variability are the variance of the powerlaw flux, and the variance of the photon index.

For our model, we assume that the photon index, $\Gamma$ is correlated with the power-law flux $F_\mathrm{PL}$, with a mean $\Gamma$ value of 2. The average absolute flux has no effect on the fractional excess variance, so we use an arbitrary value of 1. This means the final model has two parameters: the variance of the flux (in log space, $\mathrm{Var}(\log(F_\mathrm{PL}))$), and the strength of the correlation between $\log(F_\mathrm{PL})$ and $\Gamma$, $C_\Gamma$. Because we draw fluxes from a Gaussian distribution in log space, the resulting distribution of fluxes is log-normal by definition. The \textsc{xspec} table implementation also has redshift and normalization parameters, but these should be kept fixed to 0 and 1, respectively.

\begin{figure*}
    \centering
    \includegraphics[width=0.4\linewidth]{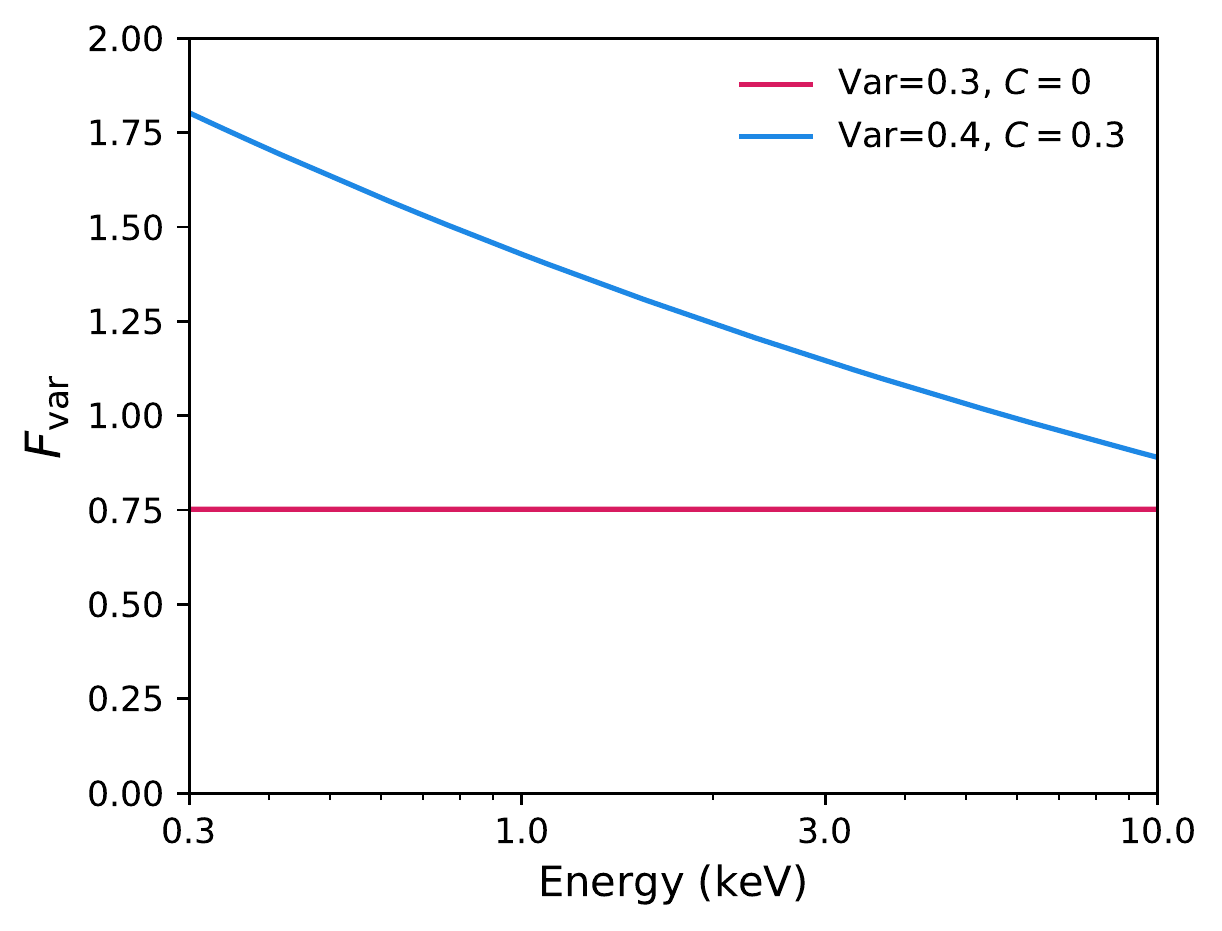}
    \includegraphics[width=0.4\linewidth]{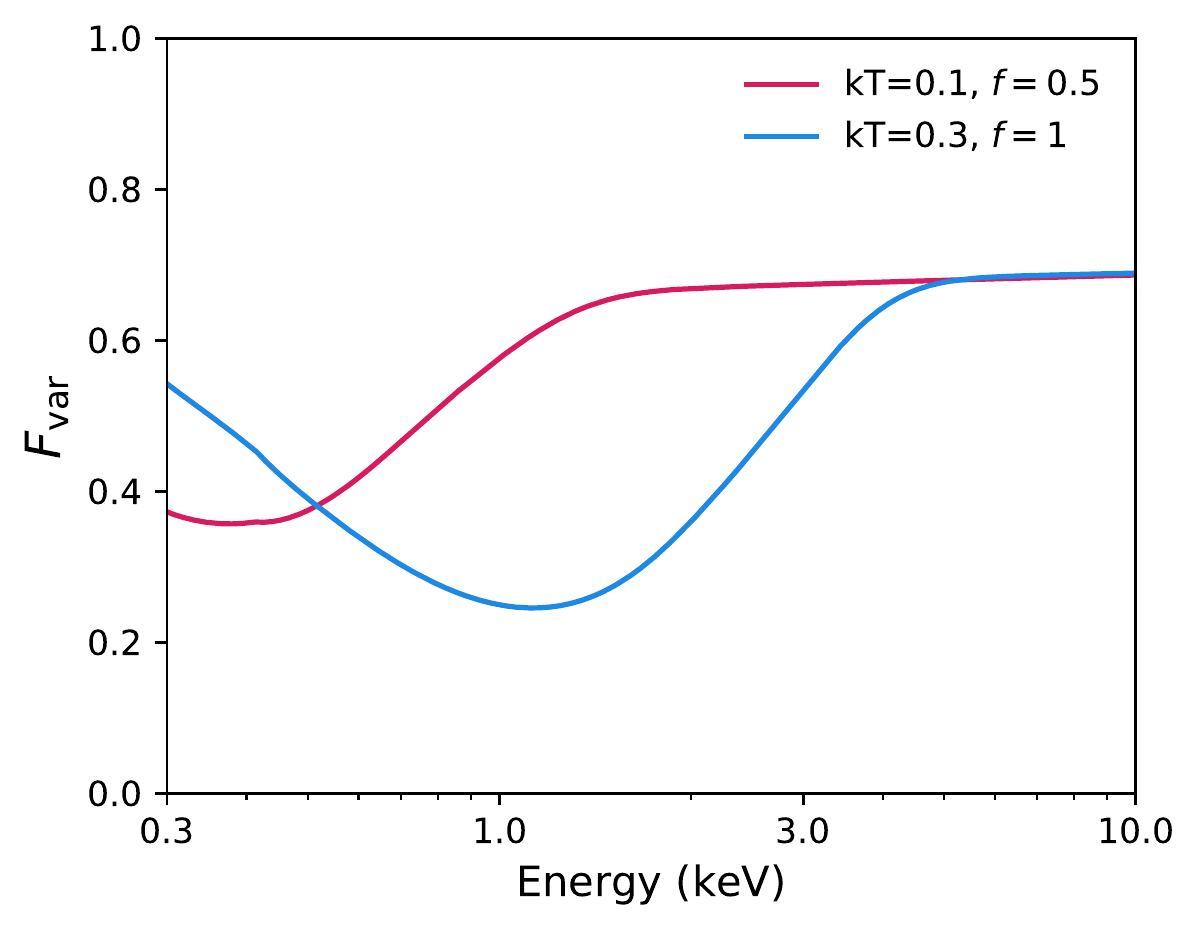}
    \includegraphics[width=0.4\linewidth]{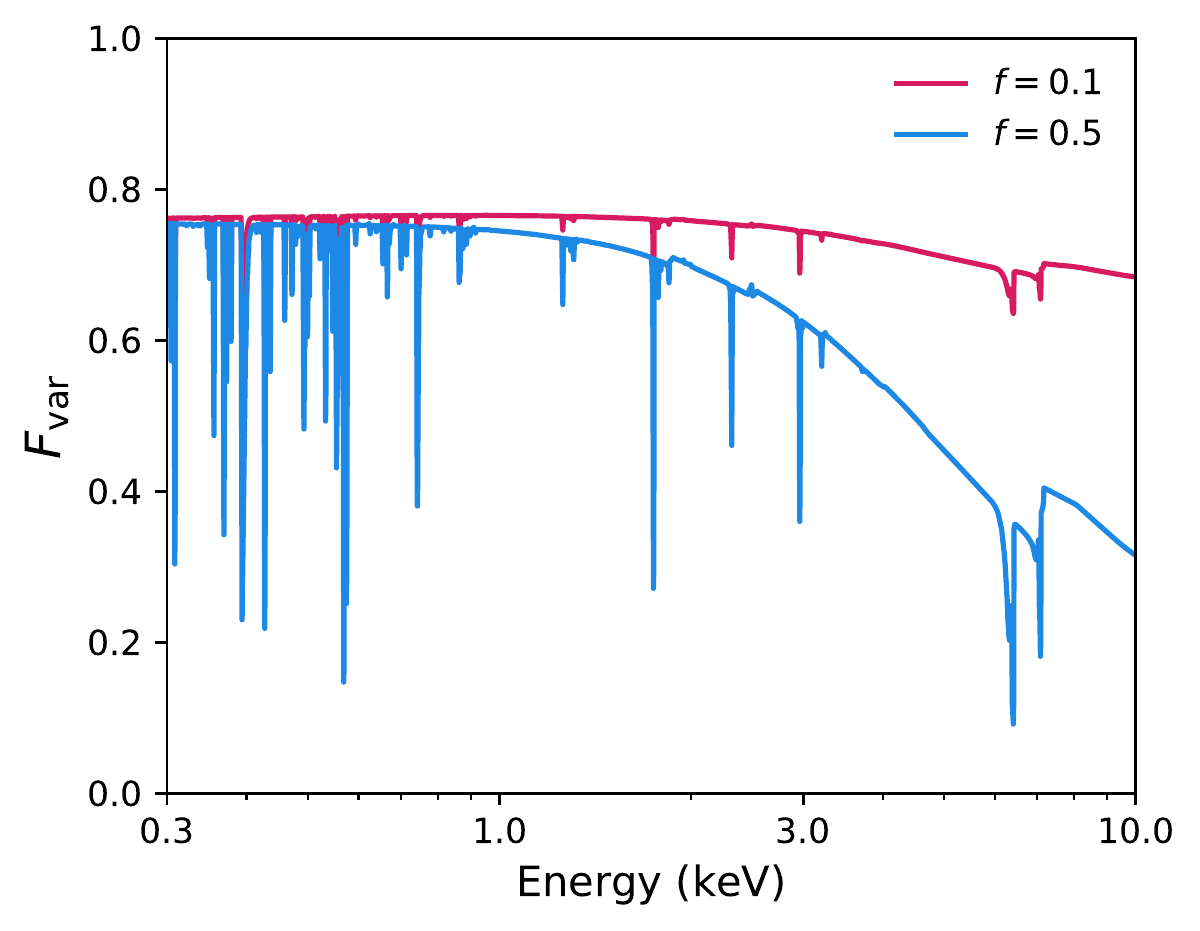}
    \includegraphics[width=0.4\linewidth]{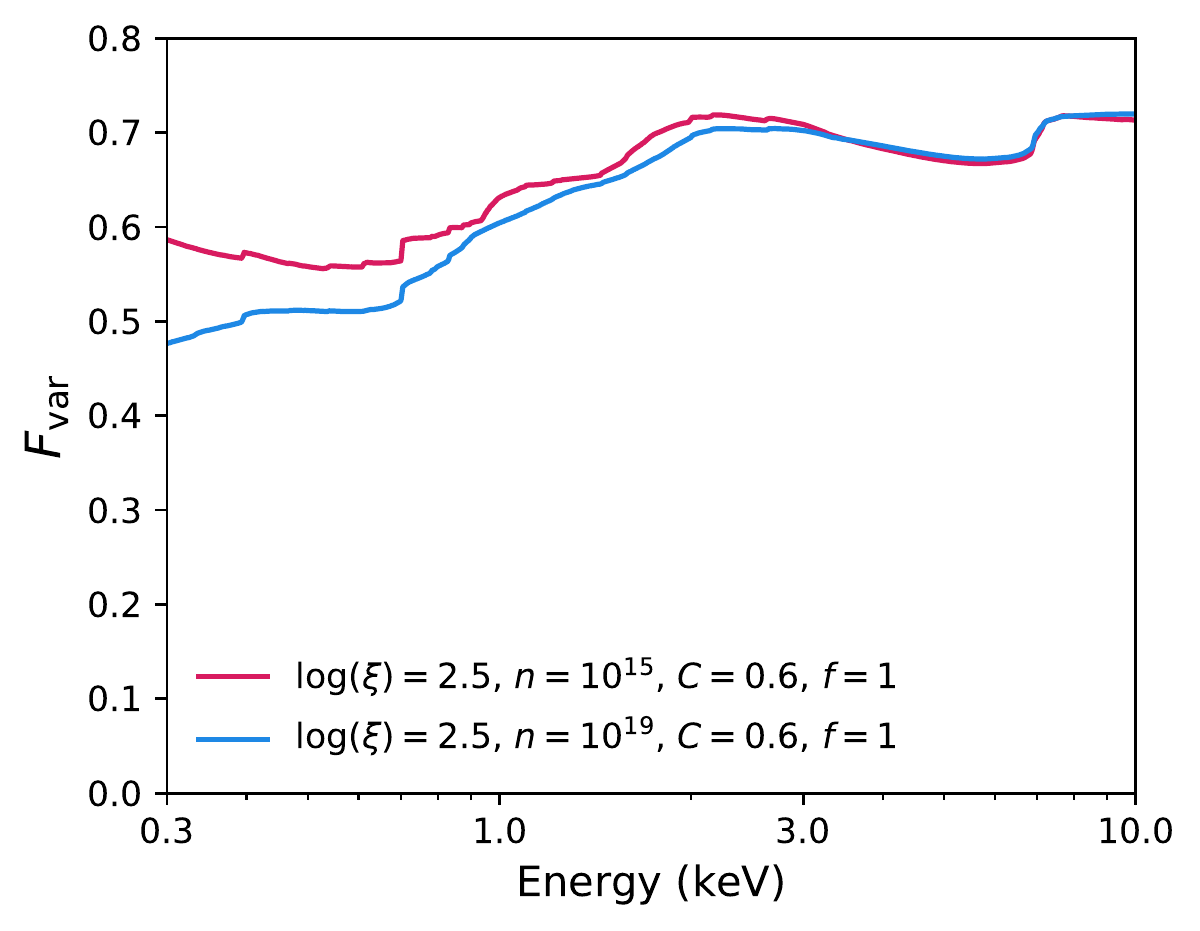}
    \includegraphics[width=0.4\linewidth]{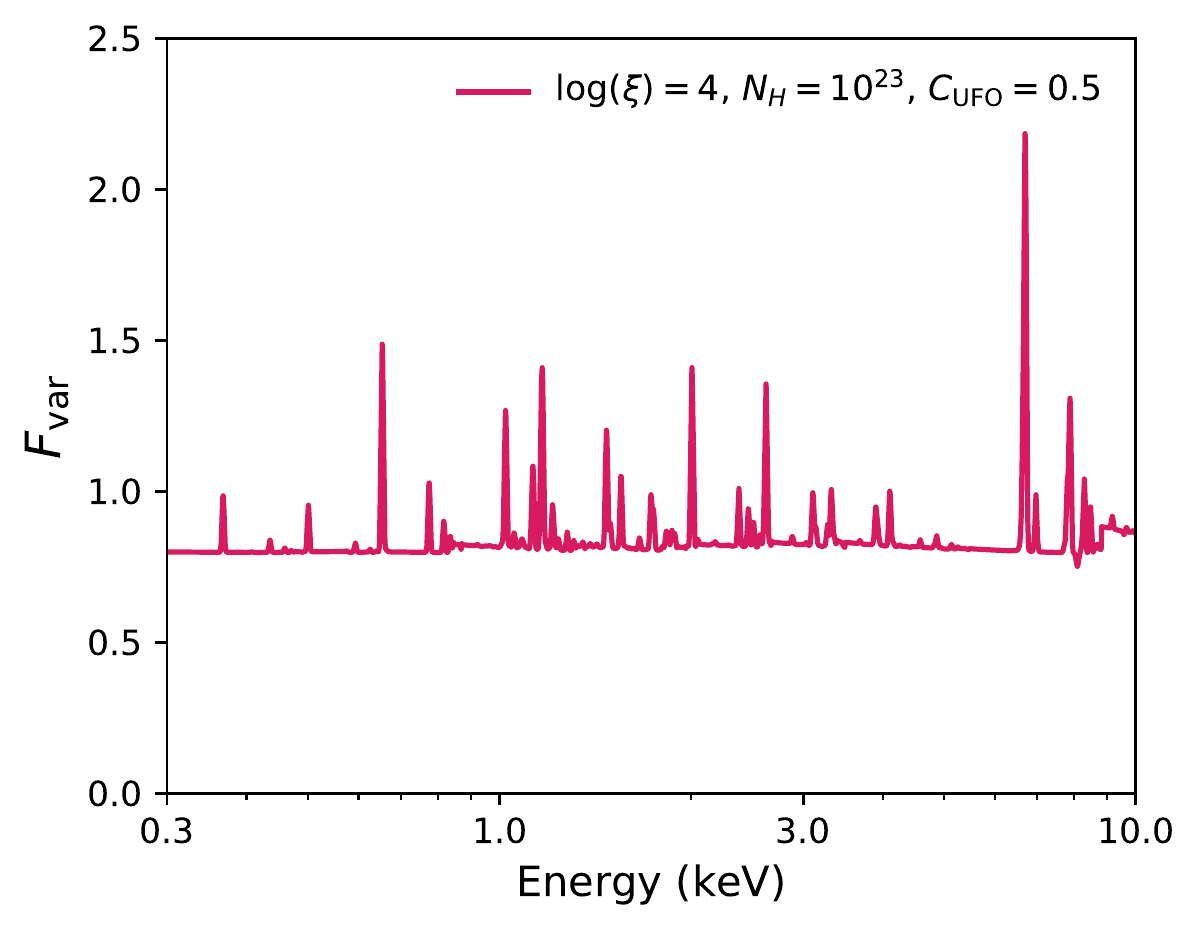}
    \caption{$F_\mathrm{var}$ model spectra for: a variable pivoting power-law (top left); variance damping from a constant black body (top right); variance damping from constant distant reflection (middle left); variance damping from relativistic reflection (middle right), where the flux is correlated with the continuum; variance enhancement from a UFO (bottom), where the ionization is correlated with the continuum.}
    \label{fig:models}
\end{figure*}

Two example model spectra are shown in Fig.~\ref{fig:models} (top left).

\subsection{Black body}

We next consider the effect of a less variable soft excess. For simplicity, we consider a constant black-body and a variable power-law. The power-law flux variance $\mathrm{Var}(\log(F_\mathrm{PL}))$ is fixed to 0.2, and the correlation with photon index $C_\Gamma$ is fixed to 0.3. The model therefore has two parameters: the ratio of the 0.3--10~keV flux in the black body to that of the average power-law, $f_\mathrm{BB}$, and the temperature of the black body, $kT$.

We calculate this grid of spectra, and then calculate the ratio of each spectrum to that of a variable power-law without a constant black body. We use these ratios as our final model: analogous to absorption in a count spectrum, \texttt{fvar\_bbdamp.fits} is a multiplicative model which lowers the variance at the energies of the black body.


\subsection{Distant reflection}
A more complex damping of the variability is produced by a constant distant reflection component, from scattering off the outer accretion disk or torus. While on long timescales distant reflection can vary, for most purposes the assumption of constant flux should suffice. 

We simulate this model in the same way as the black body model: we calculate the RMS spectra for a variable powerlaw and a constant reflection component \citep[modelled with \textsc{xillver}:][]{Garcia13}. We assume solar abundances and an ionization of 1. The model, \texttt{fvar\_xildamp.fits}, has only one parameter, the ratio of the 0.3--10 flux in the distant reflection component to that of the powerlaw, $f_\mathrm{dist}$.


Two example model spectra, applied to a variable power-law spectrum, are shown in Fig.~\ref{fig:models} (middle left).

\subsection{Relativistic reflection}
\label{sec:reflection}
The obvious next step is to simulate a relativistic reflection variance spectrum. This is more complex, as we have to consider multiple physical parameters. The most important aspect of modelling this component is the fraction of the source flux that is due to relativistic reflection as a function of energy. In particular, the strength of the soft excess, and the shape of the dip at intermediate energies are crucial. 

The relativistic Fe line has a minimal impact on the shape of the variance spectra compared to the neutral reflection case, so the exact parameters of the relativistic blurring and iron abundance assumed are less important than when fitting a count spectrum. 

For our simulations, we base the reflection model on that found for IRAS~13224-3809 by \citet{Jiang18_iras}. Using the \textsc{relxillD} model \citep{Garcia14,Garcia16}, we fix the spin to 0.98, the inner emissivity index to 5, the emissivity break radius to 6, the inclination to 50 degrees, the iron abundance to 5, and $\Gamma$ to 2.

When simulating the variance spectra, we vary only the flux of the reflection component, which we couple to the flux of the power-law via a correlation parameter $C_\mathrm{ref}$. The most important parameters for the broad-band RMS spectral shape are the density of the disc $n$, the reflection fraction $f_\mathrm{ref}$ and the ionization $\xi$. We therefore consider a range of values in each of these parameters, so our reflection model \texttt{fvar\_refdamp.fits} has 4 parameters: $n$, $C_\mathrm{ref}$, $f_\mathrm{ref}$ and $\xi$.


Two example model spectra are shown in Fig.~\ref{fig:models} (middle right).

\subsection{Outflows}
As we have shown in various papers \citep[e.g.][]{Parker17_nature, Pinto18}, the strength of absorption features from ultra-fast outflows (UFOs) is anti-correlated with the strength of the continuum. The natural consequence of this is an enhancement of the variance in energy bands corresponding to strong absorption features, manifesting as a series of peaks in variance spectra \citep{Parker17_irasvariability,Parker18_pds456}. 

This behaviour can be most simply explained by ionization of the material in the outflow by the increased X-ray flux, which will tend to weaken the absorption lines. Even if this interpretation is not correct \citep[see e.g.][]{Fukumura18}, it is consistent with the data and therefore serves as a reasonable basis from which to simulate models.

For this model, we consider a range of values of column density $N_\mathrm{H}$, mean ionization $\xi$, and correlation between the ionization and the continuum $C_\mathrm{UFO}$, again simulating a variable power-law continuum. We model the absorption using an \textsc{xspec}table model of the \textsc{xabs} model \citep{Steenbrugge03} from \textsc{spex} \citep{Kaastra96}\footnote{\textsc{xspec} implementation available here: \url{http://bit.ly/xabs_tables}}. As with the previous models, we then calculate the ratio of the variance to an equivalent model without a UFO, and use this ratio to calculate a multiplicative table model, \texttt{fvar\_ufo.fits}. An example model spectrum is shown in Fig.~\ref{fig:models} (bottom).

\section{Results}
\label{sec:results}

We now fit these models to the complex RMS spectrum from IRAS~13224-3809. The spectrum is taken from a sample of variance spectra to be presented in Igo et al. (in prep.). The spectrum is calculated from the full \xmm\ lightcurve, taking all available data and binned into 100s time bins. Full details of the data processing will be presented in Igo et al.. We transfer the spectrum to \textsc{xspec} using the \textsc{ftflx2xsp} FTOOL.

Before we perform detailed fits to RMS spectra, we must consider some general points where this differs from fitting count spectra. Firstly, RMS spectra do not usually come with associated response files. The effective area of the instrument is much less important than when fitting count spectra, as RMS spectra are fractional, but the resolution of the instrument must be taken into account for any model more complex than a power-law. There are several ways of doing this: recalculate the response matrix to correspond to the new data, construct a convolution model to account for resolution, or simply smooth the model. For this work, we simply smooth the model using an energy dependent Gaussian (\textsc{gsmooth} in \textsc{xspec}). We discuss this further in Appendix~\ref{sec:response}. To approximate the \xmm\ EPIC-pn resolution, we fix the $\sigma$ at 6~keV to 0.1~keV, and the index of the energy dependence to 0.165. 

The second factor to take into account is that the models for RMS spectra are relatively crude compared to those for count spectra. The models we calculate here do not have the same number of degrees of freedom of their count-spectra counterparts, and are therefore unlikely to be able to fit the data as well as conventional models do. This is particularly an issue for soft energies, where $F_\mathrm{var}$ can be measured to very high precision. We therefore recommend that a small systematic error be used when fitting, to accommodate uncertainties in the modelling rather than the data. The ideal amount of error to use will likely vary, but we find 1--2\% to work well: this has no effect at high energies, where the error bars are $>>2$\%, and stops the points with small errors at low energies ($<1$~keV) from dominating the fits. For the fits presented here, we use a 2\% systematic error.

Finally, we note that any constant multiplicative components (such as Galactic absorption) can be safely neglected when fitting RMS spectra, as they have no effect on the fractional variance, other than raising of lowering the signal-to-noise in affected energy bands.

\subsection{Fitting the variability of IRAS~13224-3809}
\begin{figure}
    \centering
    \includegraphics[width=\linewidth]{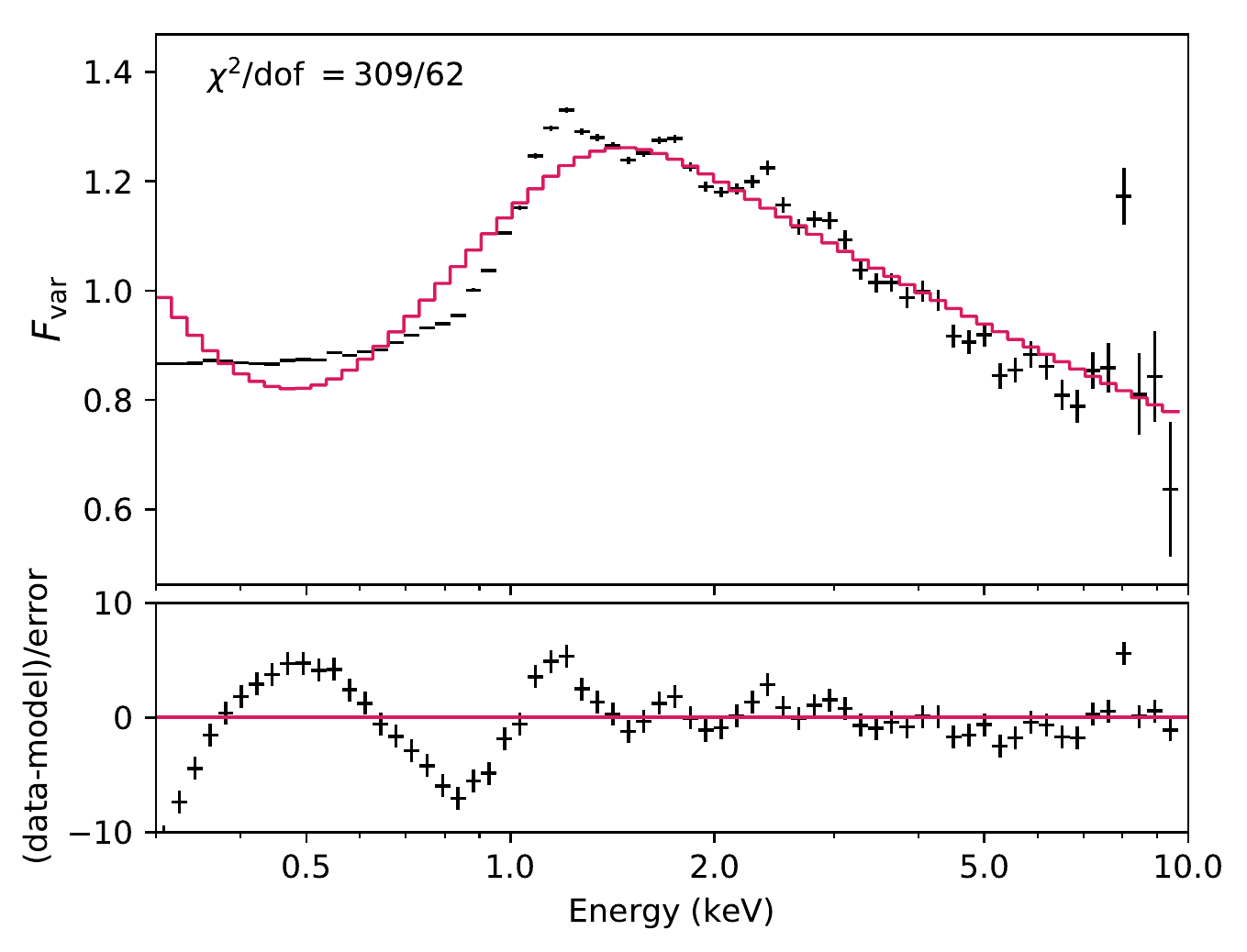}
    \includegraphics[width=\linewidth]{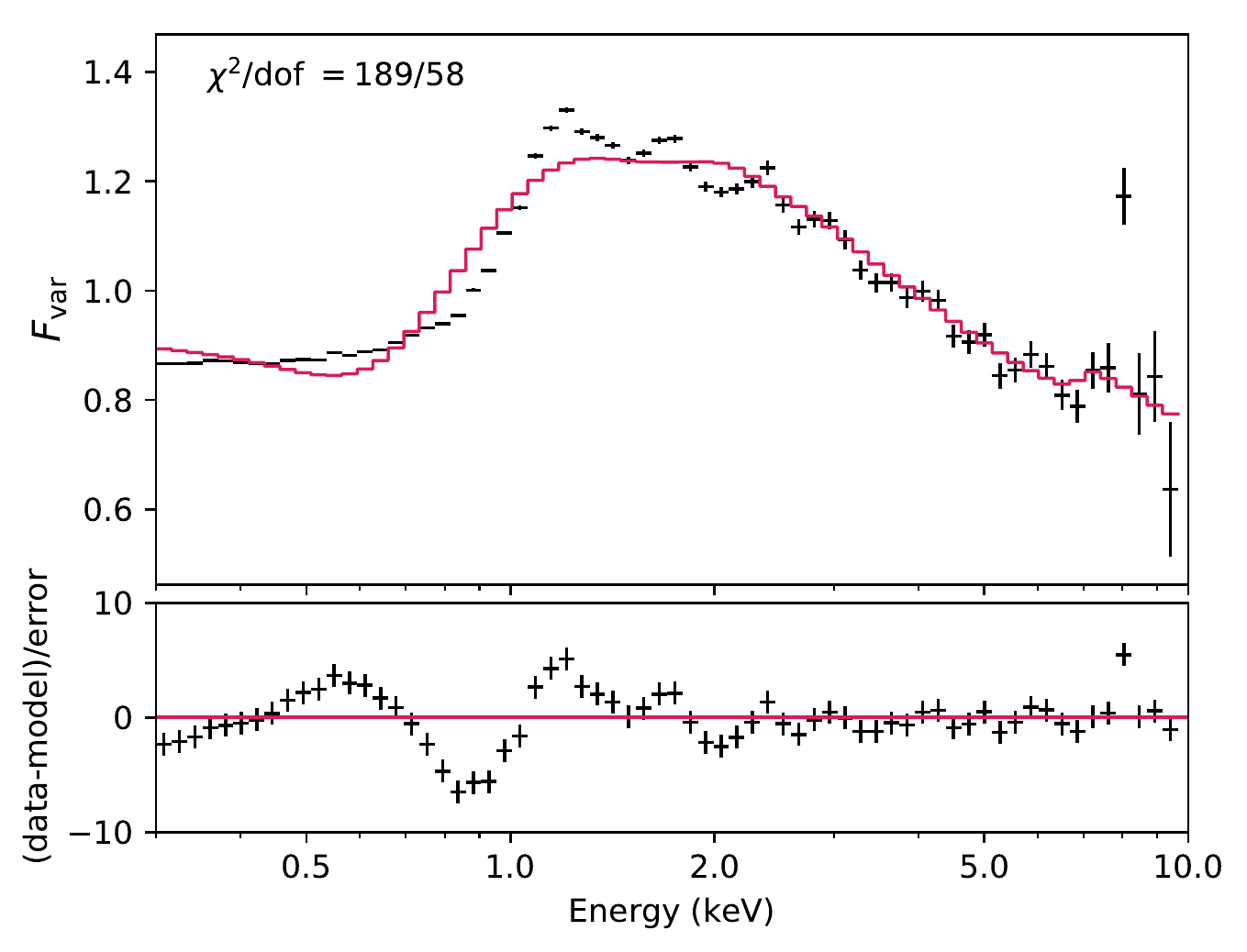}
    \includegraphics[width=\linewidth]{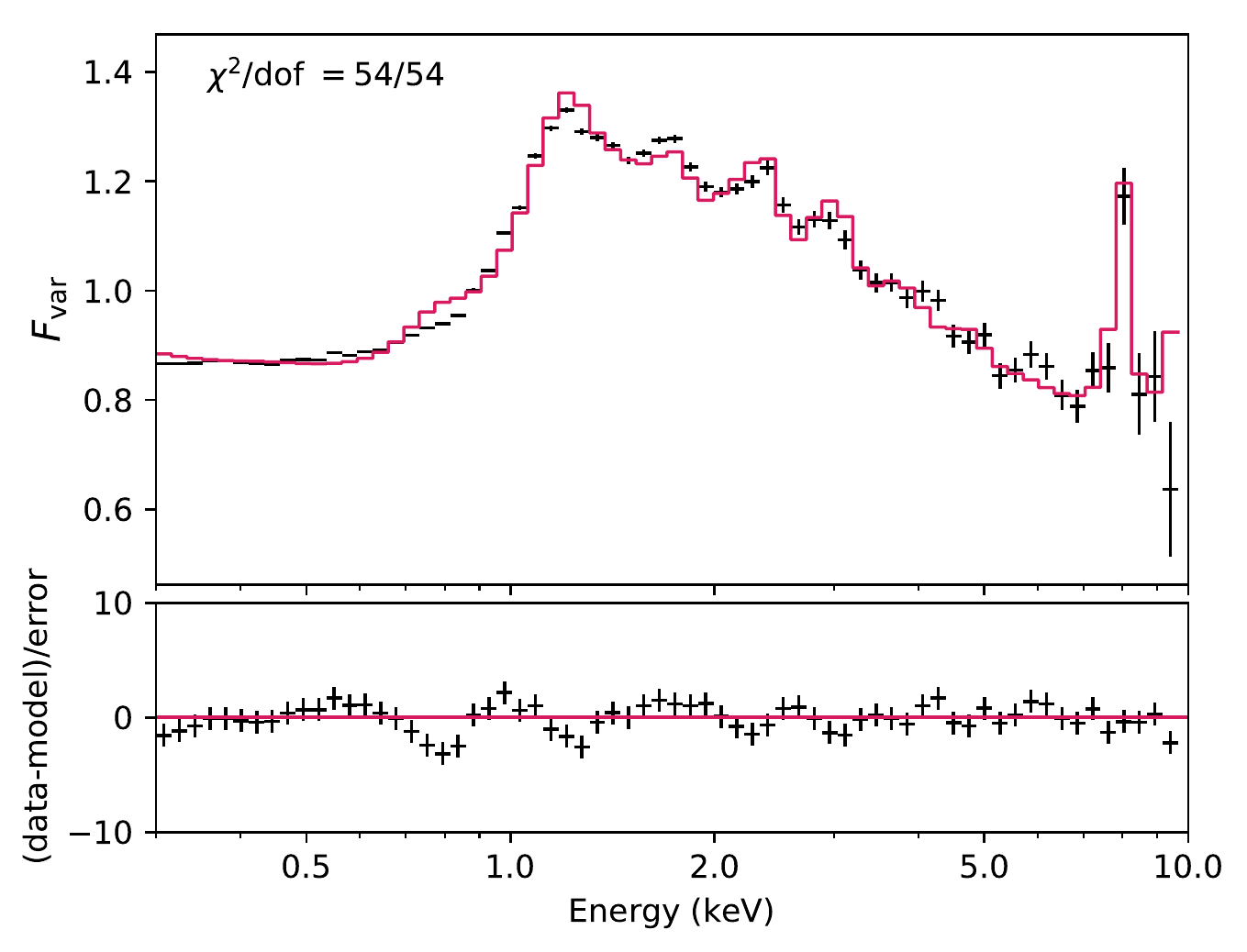}
    \caption{Intrinsic variability fits to the RMS spectrum of IRAS~13224-3809. Top: power-law variability, damped at low energies by a black-body soft excess. Middle: as top, but also damped by relativistic reflection. Bottom: as middle, but enhanced by UFO variability. In each case, the errors on the residuals include the 2\% systematic error that we add to show which bands are dominating the fit.}
    \label{fig:reffits}
\end{figure}

We start with a simple phenomenological model of a power-law damped by a black-body soft excess (\textsc{gsmooth $\times$ fvar\_bbdamp $\times$ fvar\_pow} in \textsc{xspec}). This is shown in the top panel of Fig.~\ref{fig:reffits}. It provides a simple qualitative description, but is obviously not a good fit.

To this basic model, we add a reflection component (\textsc{gsmooth $\times$ fvar\_bbdamp $\times$ fvar\_refdamp $\times$ fvar\_pow}). This clearly offers a great improvement over the simpler model ($\Delta\chi^2=120$, for 4 degrees of freedom), but is still a poor fit. In particular, it misses the peaks in the RMS spectrum due to the flux-dependent UFO absorption.

Finally, we include the UFO variance component (\textsc{gsmooth $\times$ fvar\_bbdamp $\times$ fvar\_refdamp $\times$ fvar\_ufo $\times$ fvar\_pow}). This final model provides an excellent description of the data ($\chi^2$/dof $=54/54$). The remaining residuals are due to deficiencies in modelling the soft excess, and the line ratios in the UFO not being completely accurate. The parameters of this model are given in Table~\ref{tab:refpars}.

\begin{table}
    \centering
    \caption{Best fit model parameters for the intrinsic variability model shown in the bottom panel of Fig.~\ref{fig:reffits}.}
    \begin{tabular}{l c c r}
    \hline
    \hline
    Component & Parameter & Value & Unit\\
    \hline
    \textsc{fvar\_pow}   &   Var & $0.40_{-0.01}^{+0.03}$  &\\
                        &   $C_\Gamma$  & $0.31\pm0.004$  &\\
    \textsc{fvar\_bbdamp}&   kT  & $0.096_{-0.006}^{+0.005}$  &   keV\\
                        &   $f_\mathrm{BB}$ & $0.13\pm0.03$  &\\
    \textsc{fvar\_refdamp}& $\log(n)$ & $17.3_{-0.7}^{+0.5}$  &   cm$^{-3}$\\
                        &   $\log(\xi)$   & $>2.9$  & erg~cm~s$^{-1}$\\
                        &   $f_\mathrm{ref}$&  $1.1_{-0.5}^{+1.0}$ &\\
                        &   $C_\mathrm{ref}$& $0.59_{-0.12}^{+0.06}$  &\\
    \textsc{fvar\_ufo}  &   $\log(\xi)$   & $3.901\pm0.005$  &erg~cm~s$^{-1}$\\
                        &      $n_\mathrm{H}$ & $9.6_{-1.6}^{+0.4}\times10^{22}$ & cm$^{-2}$\\
                        &   $C_\mathrm{UFO}$& $0.76_{-0.01}^{+0.02}$  &\\
                        &   $v$ & $-0.232\pm0.004$  &    $c$\\
    \hline
    \hline
    \end{tabular}
    \label{tab:refpars}
\end{table}

\section{Discussion}

We have presented several models for fitting the excess variance spectra of AGN, and demonstrated an example case modelling the complex RMS spectrum of IRAS~13224-3809. These models, and the code used to generate them, are publicly available here: \url{http://bit.ly/fvar_models}.

One promising application of this technique is the detection of highly ionised outflows. We showed in \citet{Parker17_irasvariability, Parker18_pds456} that UFOs produce strong features in variance spectra, but for borderline cases there was no method to quantify the significance of an outflow seen in the RMS spectrum. With the new models we have presented, it is now trivial to fit such a spectrum and find the statistical improvement offered by including a UFO.

From our fits, it is clear that the intrinsic variability model, comprised of a variable power-law, a less variable reflection component and black body, and a UFO which enhances the variability, offers an excellent description of the variance spectrum of IRAS~13224-3809. The black body component, which is also required in the count spectrum \citep[e.g.][]{Chiang15, Jiang18_iras}, gives a significant improvement to the fit of the soft excess, although most of it is still described by reflection. It is not obvious whether that is because some non-reflection contribution is required, or because our reflection model is not yet sophisticated enough to fit the soft excess properly.

The RMS spectrum of IRAS~13224-3809 was also examined by \citet{Ponti10} and \citet{Fabian13_iras}. These papers predated the 2016 1.5~Ms \xmm\ campaign (PI Fabian), so the quality of the data was significantly lower. As such, the UFO lines were not resolved. In particular, the Fe line at 8~keV was interpreted as the power-law fraction increasing after the blue wing of the relativistic Fe~K line. As shown in Fig.~\ref{fig:reffits} and by \citet{Yamasaki16}, the drop in variance caused by the relativistic Fe line is small in this case, and cannot easily explain the size of the 7~keV dip. With the higher quality spectrum now available, it is clear that this dip is not a real feature, rather there is a peak at 8~keV.

\citet{Yamasaki16} were able to successfully fit the RMS spectra of IRAS~13224-3809 with a partial-covering model. However, we consider this result unreliable, for several reasons. Firstly, the ad-hoc nature of the model used to fit the spectra, with an artificial edge at 1.2~keV, means that their model is not physical. \citeauthor{Yamasaki16} attribute this edge to Fe~L, but this requires an extreme blueshift as the rest energy of Fe~L is 0.7--0.8~keV.

Secondly, their RMS spectra are not simulated from first principles: instead, the model is fit to the data in time-slices and an RMS spectrum simulated from the resulting time-resolved modelling. This all but guarantees that the simulated RMS spectrum will match that of the data. Our intrinsic variability dominated model is generated from first principles, and gives an excellent description of the data without fine-tuning. 

Additionally, the RMS spectra used by \citeauthor{Yamasaki16} were for individual observations, and were therefore of much lower quality than the total RMS spectrum shown here. In particular, the UFO lines were not resolved. This lead both these authors and \citet{Ponti10} to conclude that the variability increased sharply above 7~keV and never recovered. With higher quality data, it is clear that this is not the case.

Finally, we note that all such absorption models predict an abundance of absorption lines in the soft band which would be easily detectable with grating spectra. The standard argument to avoid this is that the lines must be smeared by velocity broadening so they are not observable, however this is not self-consistent. In order to produce variability of the scale observed on the timescales observed, both emission and absorption regions must be compact, otherwise the covering fractions would not be able to change rapidly. Therefore, our line of sight can only be intercepting a small number of absorbing clouds, which will have well defined velocities. In this scenario, absorption lines in the RGS are unavoidable. No such lines are present \citep[e.g.][]{Parker17_nature,Pinto18}.

As a general rule, absorption models struggle to explain the fundamentals of AGN variability. They do not naturally produce the RMS-flux relation over a wide range of frequencies, or a log-normal flux distribution without fine-tuning, and offer no explanation for the wide range of frequencies that AGN vary on \citep[the power spectrum shown by][for example, shows that the AGN has power over six decades in frequency]{Alston19}.



It has been long established that reflected emission is less variable than the continuum, and this is frequently visible in variance spectra \citep[e.g.][]{Vaughan04}, which lead to the development of the light-bending model \citep{Miniutti03, Miniutti04}. In this scenario, the continuum variability is at least partly due to changes in the geometry (primarily the height) of the X-ray corona, such that the fraction of coronal photons escaping to infinity or falling into the black hole changes drastically, while only minor changes in the fraction hitting the disk occur. This is in slight tension with the requirement that the reflection flux be correlated with the continuum, in order for reverberation lags to be detected \citep[e.g.][]{Fabian09}.
While previous simulations of variance spectra have simply assumed a constant reflection component \citep[as in][]{Ponti10}, the reflection model we present in Section~\ref{sec:reflection} allows the strength of the relation between continuum and reflected flux to be quantified: for IRAS~13224-3809, we find a correlation of $\sim0.6$ between the two. This shows that the reflection flux does track the continuum, but not at a 1:1 ratio. 

The discrepancy most likely arises because RMS spectra are generally calculated over a wide range of frequencies, with the lowest frequencies dominating because they contain the most power, whereas the reverberation lags are typically found at high frequencies. These results are therefore perfectly consistent, requiring only that the response of the reflection to the continuum be stronger at high frequencies. This is a reasonable assumption, as the highest frequency variability is likely intrinsic to the corona and should provoke the strongest response from the reflected component.

This technique can be used to directly probe the strength of the reflection/continuum correlation as a function of frequency: in Alston et al. (submitted), we show that the reflection tracks the continuum more closely at higher frequencies. This implies that at low frequency geometry changes dominate the variability, while at high frequency the variability is intrinsic to the corona. In future work, we will use this to examine the disc-corona interaction in detail. We also note that the continuum/reflection correlation will include information about the lag between the components: a longer delay between the continuum and reflected flux will cause a weaker correlation between the two at high frequencies. More detailed modelling is needed to see how these two effects interact.



\section{Conclusions}

We have presented \textsc{xspec} models for fitting variance spectra of AGN dominated by intrinsic variability. Our models are derived from Monte-Carlo simulations using basic assumptions about AGN variability, and give an excellent fit to the complex RMS spectrum of IRAS~13224-3809.

The fit to the IRAS~13224-3809 spectrum shows that a combination of a variable power-law, moderately correlated reflection, and an outflow where the ionization responds to the X-ray emission fully describes the observed variance. The model is generally consistent with that used to fit the count spectrum, and fitting the variance allows us to constrain relations between different components that are not available through conventional spectroscopy.

We intend to improve these models, and make more similar models in future, so that RMS spectra can routinely be fit alongside conventional spectra.


\section*{Acknowledgements}
MLP is supported by European Space Agency (ESA) Research Fellowships. ACF and WNA acknowledge support from ERC Advanced Grant 340442. Based on observations obtained with XMM-Newton, an ESA science mission with instruments and contributions directly funded by ESA Member States and NASA. We thank the anonymous referee for their constructive feedback.




\bibliographystyle{mnras}
\bibliography{bibliography} 

\begin{thebibliography}{}
\makeatletter
\relax
\def\mn@urlcharsother{\let\do\@makeother \do\$\do\&\do\#\do\^\do\_\do\%\do\~}
\def\mn@doi{\begingroup\mn@urlcharsother \@ifnextchar [ {\mn@doi@}
  {\mn@doi@[]}}
\def\mn@doi@[#1]#2{\def\@tempa{#1}\ifx\@tempa\@empty \href
  {http://dx.doi.org/#2} {doi:#2}\else \href {http://dx.doi.org/#2} {#1}\fi
  \endgroup}
\def\mn@eprint#1#2{\mn@eprint@#1:#2::\@nil}
\def\mn@eprint@arXiv#1{\href {http://arxiv.org/abs/#1} {{\tt arXiv:#1}}}
\def\mn@eprint@dblp#1{\href {http://dblp.uni-trier.de/rec/bibtex/#1.xml}
  {dblp:#1}}
\def\mn@eprint@#1:#2:#3:#4\@nil{\def\@tempa {#1}\def\@tempb {#2}\def\@tempc
  {#3}\ifx \@tempc \@empty \let \@tempc \@tempb \let \@tempb \@tempa \fi \ifx
  \@tempb \@empty \def\@tempb {arXiv}\fi \@ifundefined
  {mn@eprint@\@tempb}{\@tempb:\@tempc}{\expandafter \expandafter \csname
  mn@eprint@\@tempb\endcsname \expandafter{\@tempc}}}

\bibitem[\protect\citeauthoryear{{Alston} et~al.,}{{Alston}
  et~al.}{2019}]{Alston19}
{Alston} W.~N.,  et~al., 2019, \mn@doi [\mnras] {10.1093/mnras/sty2527}, \href
  {https://ui.adsabs.harvard.edu/abs/2019MNRAS.482.2088A} {482, 2088}

\bibitem[\protect\citeauthoryear{{Ar{\'e}valo} \& {Uttley}}{{Ar{\'e}valo} \&
  {Uttley}}{2006}]{Arevalo06}
{Ar{\'e}valo} P.,  {Uttley} P.,  2006, \mn@doi [\mnras]
  {10.1111/j.1365-2966.2006.09989.x}, \href
  {https://ui.adsabs.harvard.edu/abs/2006MNRAS.367..801A} {367, 801}

\bibitem[\protect\citeauthoryear{{Arnaud}}{{Arnaud}}{1996}]{Arnaud96}
{Arnaud} K.~A.,  1996, in {Jacoby} G.~H.,  {Barnes} J.,  eds,  Astronomical
  Society of the Pacific Conference Series Vol. 101, Astronomical Data Analysis
  Software and Systems V. p.~17

\bibitem[\protect\citeauthoryear{{Brenneman} et~al.,}{{Brenneman}
  et~al.}{2014}]{Brenneman14}
{Brenneman} L.~W.,  et~al., 2014, \mn@doi [\apj] {10.1088/0004-637X/788/1/61},
  \href {http://adsabs.harvard.edu/abs/2014ApJ...788...61B} {788, 61}

\bibitem[\protect\citeauthoryear{{Chiang}, {Walton}, {Fabian}, {Wilkins}  \&
  {Gallo}}{{Chiang} et~al.}{2015}]{Chiang15}
{Chiang} C.-Y.,  {Walton} D.~J.,  {Fabian} A.~C.,  {Wilkins} D.~R.,   {Gallo}
  L.~C.,  2015, \mn@doi [\mnras] {10.1093/mnras/stu2087}, \href
  {http://adsabs.harvard.edu/abs/2015MNRAS.446..759C} {446, 759}

\bibitem[\protect\citeauthoryear{{Edelson}, {Turner}, {Pounds}, {Vaughan},
  {Markowitz}, {Marshall}, {Dobbie}  \& {Warwick}}{{Edelson}
  et~al.}{2002}]{Edelson02}
{Edelson} R.,  {Turner} T.~J.,  {Pounds} K.,  {Vaughan} S.,  {Markowitz} A.,
  {Marshall} H.,  {Dobbie} P.,   {Warwick} R.,  2002, \mn@doi [\apj]
  {10.1086/323779}, \href {http://adsabs.harvard.edu/abs/2002ApJ...568..610E}
  {568, 610}

\bibitem[\protect\citeauthoryear{{Fabian} et~al.,}{{Fabian}
  et~al.}{2009}]{Fabian09}
{Fabian} A.~C.,  et~al., 2009, \mn@doi [\nat] {10.1038/nature08007}, \href
  {http://adsabs.harvard.edu/abs/2009Natur.459..540F} {459, 540}

\bibitem[\protect\citeauthoryear{{Fabian} et~al.,}{{Fabian}
  et~al.}{2013}]{Fabian13_iras}
{Fabian} A.~C.,  et~al., 2013, \mn@doi [\mnras] {10.1093/mnras/sts504}, \href
  {http://adsabs.harvard.edu/abs/2013MNRAS.429.2917F} {429, 2917}

\bibitem[\protect\citeauthoryear{{Fukumura}, {Kazanas}, {Shrader}, {Behar},
  {Tombesi}  \& {Contopoulos}}{{Fukumura} et~al.}{2018}]{Fukumura18}
{Fukumura} K.,  {Kazanas} D.,  {Shrader} C.,  {Behar} E.,  {Tombesi} F.,
  {Contopoulos} I.,  2018, \mn@doi [\apjl] {10.3847/2041-8213/aadd10}, \href
  {https://ui.adsabs.harvard.edu/abs/2018ApJ...864L..27F} {864, L27}

\bibitem[\protect\citeauthoryear{{Garc{\'{\i}}a}, {Dauser}, {Reynolds},
  {Kallman}, {McClintock}, {Wilms}  \& {Eikmann}}{{Garc{\'{\i}}a}
  et~al.}{2013}]{Garcia13}
{Garc{\'{\i}}a} J.,  {Dauser} T.,  {Reynolds} C.~S.,  {Kallman} T.~R.,
  {McClintock} J.~E.,  {Wilms} J.,   {Eikmann} W.,  2013, \mn@doi [\apj]
  {10.1088/0004-637X/768/2/146}, \href
  {http://adsabs.harvard.edu/abs/2013ApJ...768..146G} {768, 146}

\bibitem[\protect\citeauthoryear{{Garc{\'{\i}}a} et~al.,}{{Garc{\'{\i}}a}
  et~al.}{2014}]{Garcia14}
{Garc{\'{\i}}a} J.,  et~al., 2014, \mn@doi [\apj] {10.1088/0004-637X/782/2/76},
  \href {http://adsabs.harvard.edu/abs/2014ApJ...782...76G} {782, 76}

\bibitem[\protect\citeauthoryear{{Garc{\'\i}a}, {Fabian}, {Kallman}, {Dauser},
  {Parker}, {McClintock}, {Steiner}  \& {Wilms}}{{Garc{\'\i}a}
  et~al.}{2016}]{Garcia16}
{Garc{\'\i}a} J.~A.,  {Fabian} A.~C.,  {Kallman} T.~R.,  {Dauser} T.,  {Parker}
  M.~L.,  {McClintock} J.~E.,  {Steiner} J.~F.,   {Wilms} J.,  2016, \mn@doi
  [\mnras] {10.1093/mnras/stw1696}, \href
  {https://ui.adsabs.harvard.edu/abs/2016MNRAS.462..751G} {462, 751}

\bibitem[\protect\citeauthoryear{{Hogg} \& {Reynolds}}{{Hogg} \&
  {Reynolds}}{2016}]{Hogg16}
{Hogg} J.~D.,  {Reynolds} C.~S.,  2016, \mn@doi [\apj]
  {10.3847/0004-637X/826/1/40}, \href
  {https://ui.adsabs.harvard.edu/abs/2016ApJ...826...40H} {826, 40}

\bibitem[\protect\citeauthoryear{{Jiang} et~al.,}{{Jiang}
  et~al.}{2018}]{Jiang18_iras}
{Jiang} J.,  et~al., 2018, \mn@doi [\mnras] {10.1093/mnras/sty836}, \href
  {https://ui.adsabs.harvard.edu/abs/2018MNRAS.477.3711J} {477, 3711}

\bibitem[\protect\citeauthoryear{{Jin}, {Done}, {Middleton}  \& {Ward}}{{Jin}
  et~al.}{2013}]{Jin13}
{Jin} C.,  {Done} C.,  {Middleton} M.,   {Ward} M.,  2013, \mn@doi [\mnras]
  {10.1093/mnras/stt1801}, \href
  {https://ui.adsabs.harvard.edu/abs/2013MNRAS.436.3173J} {436, 3173}

\bibitem[\protect\citeauthoryear{{Kaastra}, {Mewe}  \&
  {Nieuwenhuijzen}}{{Kaastra} et~al.}{1996}]{Kaastra96}
{Kaastra} J.~S.,  {Mewe} R.,   {Nieuwenhuijzen} H.,  1996, in {Yamashita} K.,
  {Watanabe} T.,  eds, UV and X-ray Spectroscopy of Astrophysical and
  Laboratory Plasmas. pp 411--414

\bibitem[\protect\citeauthoryear{{King}, {Pringle}, {West}  \& {Livio}}{{King}
  et~al.}{2004}]{King04}
{King} A.~R.,  {Pringle} J.~E.,  {West} R.~G.,   {Livio} M.,  2004, \mn@doi
  [\mnras] {10.1111/j.1365-2966.2004.07322.x}, \href
  {https://ui.adsabs.harvard.edu/abs/2004MNRAS.348..111K} {348, 111}

\bibitem[\protect\citeauthoryear{{Lyubarskii}}{{Lyubarskii}}{1997}]{Lyubarskii97}
{Lyubarskii} Y.~E.,  1997, \mn@doi [\mnras] {10.1093/mnras/292.3.679}, \href
  {https://ui.adsabs.harvard.edu/abs/1997MNRAS.292..679L} {292, 679}

\bibitem[\protect\citeauthoryear{{Mallick}, {Dewangan}, {McHardy}  \&
  {Pahari}}{{Mallick} et~al.}{2017}]{Mallick17}
{Mallick} L.,  {Dewangan} G.~C.,  {McHardy} I.~M.,   {Pahari} M.,  2017,
  \mn@doi [\mnras] {10.1093/mnras/stx1960}, \href
  {https://ui.adsabs.harvard.edu/abs/2017MNRAS.472..174M} {472, 174}

\bibitem[\protect\citeauthoryear{{Matzeu}, {Reeves}, {Nardini}, {Braito},
  {Costa}, {Tombesi}  \& {Gofford}}{{Matzeu} et~al.}{2016}]{Matzeu16}
{Matzeu} G.~A.,  {Reeves} J.~N.,  {Nardini} E.,  {Braito} V.,  {Costa} M.~T.,
  {Tombesi} F.,   {Gofford} J.,  2016, \mn@doi [\mnras] {10.1093/mnras/stw354},
  \href {http://adsabs.harvard.edu/abs/2016MNRAS.458.1311M} {458, 1311}

\bibitem[\protect\citeauthoryear{{Middleton}, {Done}, {Ward}, {Gierli{\'n}ski}
  \& {Schurch}}{{Middleton} et~al.}{2009}]{Middleton09}
{Middleton} M.,  {Done} C.,  {Ward} M.,  {Gierli{\'n}ski} M.,   {Schurch} N.,
  2009, \mn@doi [\mnras] {10.1111/j.1365-2966.2008.14255.x}, \href
  {http://adsabs.harvard.edu/abs/2009MNRAS.394..250M} {394, 250}

\bibitem[\protect\citeauthoryear{{Miniutti} \& {Fabian}}{{Miniutti} \&
  {Fabian}}{2004}]{Miniutti04}
{Miniutti} G.,  {Fabian} A.~C.,  2004, \mn@doi [\mnras]
  {10.1111/j.1365-2966.2004.07611.x}, \href
  {http://adsabs.harvard.edu/abs/2004MNRAS.349.1435M} {349, 1435}

\bibitem[\protect\citeauthoryear{{Miniutti}, {Fabian}, {Goyder}  \&
  {Lasenby}}{{Miniutti} et~al.}{2003}]{Miniutti03}
{Miniutti} G.,  {Fabian} A.~C.,  {Goyder} R.,   {Lasenby} A.~N.,  2003, \mn@doi
  [\mnras] {10.1046/j.1365-8711.2003.06988.x}, \href
  {http://adsabs.harvard.edu/abs/2003MNRAS.344L..22M} {344, L22}

\bibitem[\protect\citeauthoryear{{Parker} et~al.,}{{Parker}
  et~al.}{2017a}]{Parker17_irasvariability}
{Parker} M.~L.,  et~al., 2017a, \mn@doi [\mnras] {10.1093/mnras/stx945}, \href
  {http://adsabs.harvard.edu/abs/2017MNRAS.469.1553P} {469, 1553}

\bibitem[\protect\citeauthoryear{{Parker} et~al.,}{{Parker}
  et~al.}{2017b}]{Parker17_nature}
{Parker} M.~L.,  et~al., 2017b, \mn@doi [\nat] {10.1038/nature21385}, \href
  {http://adsabs.harvard.edu/abs/2017Natur.543...83P} {543, 83}

\bibitem[\protect\citeauthoryear{{Parker}, {Reeves}, {Matzeu}, {Buisson}  \&
  {Fabian}}{{Parker} et~al.}{2018}]{Parker18_pds456}
{Parker} M.~L.,  {Reeves} J.~N.,  {Matzeu} G.~A.,  {Buisson} D.~J.~K.,
  {Fabian} A.~C.,  2018, \mn@doi [\mnras] {10.1093/mnras/stx2803}, \href
  {http://adsabs.harvard.edu/abs/2018MNRAS.474..108P} {474, 108}

\bibitem[\protect\citeauthoryear{{Pinto} et~al.,}{{Pinto}
  et~al.}{2018}]{Pinto18}
{Pinto} C.,  et~al., 2018, \mn@doi [\mnras] {10.1093/mnras/sty231}, \href
  {https://ui.adsabs.harvard.edu/abs/2018MNRAS.476.1021P} {476, 1021}

\bibitem[\protect\citeauthoryear{{Ponti}, {Miniutti}, {Cappi}, {Maraschi},
  {Fabian}  \& {Iwasawa}}{{Ponti} et~al.}{2006}]{Ponti06}
{Ponti} G.,  {Miniutti} G.,  {Cappi} M.,  {Maraschi} L.,  {Fabian} A.~C.,
  {Iwasawa} K.,  2006, \mn@doi [\mnras] {10.1111/j.1365-2966.2006.10165.x},
  \href {http://adsabs.harvard.edu/abs/2006MNRAS.368..903P} {368, 903}

\bibitem[\protect\citeauthoryear{{Ponti} et~al.,}{{Ponti}
  et~al.}{2010}]{Ponti10}
{Ponti} G.,  et~al., 2010, \mn@doi [\mnras] {10.1111/j.1365-2966.2010.16852.x},
  \href {https://ui.adsabs.harvard.edu/abs/2010MNRAS.406.2591P} {406, 2591}

\bibitem[\protect\citeauthoryear{{Steenbrugge} et~al.,}{{Steenbrugge}
  et~al.}{2003}]{Steenbrugge03}
{Steenbrugge} K.~C.,  et~al., 2003, \mn@doi [\aap]
  {10.1051/0004-6361:20031021}, \href
  {https://ui.adsabs.harvard.edu/abs/2003A&A...408..921S} {408, 921}

\bibitem[\protect\citeauthoryear{{Uttley} \& {McHardy}}{{Uttley} \&
  {McHardy}}{2001}]{Uttley01}
{Uttley} P.,  {McHardy} I.~M.,  2001, \mn@doi [\mnras]
  {10.1046/j.1365-8711.2001.04496.x}, \href
  {https://ui.adsabs.harvard.edu/abs/2001MNRAS.323L..26U} {323, L26}

\bibitem[\protect\citeauthoryear{{Uttley}, {McHardy}  \& {Vaughan}}{{Uttley}
  et~al.}{2005}]{Uttley05}
{Uttley} P.,  {McHardy} I.~M.,   {Vaughan} S.,  2005, \mn@doi [\mnras]
  {10.1111/j.1365-2966.2005.08886.x}, \href
  {https://ui.adsabs.harvard.edu/abs/2005MNRAS.359..345U} {359, 345}

\bibitem[\protect\citeauthoryear{{Vaughan} \& {Fabian}}{{Vaughan} \&
  {Fabian}}{2004}]{Vaughan04}
{Vaughan} S.,  {Fabian} A.~C.,  2004, \mn@doi [\mnras]
  {10.1111/j.1365-2966.2004.07456.x}, \href
  {http://adsabs.harvard.edu/abs/2004MNRAS.348.1415V} {348, 1415}

\bibitem[\protect\citeauthoryear{{Vaughan}, {Edelson}, {Warwick}  \&
  {Uttley}}{{Vaughan} et~al.}{2003}]{Vaughan03_variability}
{Vaughan} S.,  {Edelson} R.,  {Warwick} R.~S.,   {Uttley} P.,  2003, \mn@doi
  [\mnras] {10.1046/j.1365-2966.2003.07042.x}, \href
  {http://adsabs.harvard.edu/abs/2003MNRAS.345.1271V} {345, 1271}

\bibitem[\protect\citeauthoryear{{Yamasaki}, {Mizumoto}, {Ebisawa}  \&
  {Sameshima}}{{Yamasaki} et~al.}{2016}]{Yamasaki16}
{Yamasaki} H.,  {Mizumoto} M.,  {Ebisawa} K.,   {Sameshima} H.,  2016, \mn@doi
  [\pasj] {10.1093/pasj/psw070}, \href
  {https://ui.adsabs.harvard.edu/abs/2016PASJ...68...80Y} {68, 80}

\makeatother
\end{thebibliography}


\appendix
\section{Model parameter table}
\label{sec:tables}

\begin{table}
    \centering
    \caption{Parameters of the variance models presented in section~\ref{sec:models}}
    \begin{tabular}{l c c c r r}
    \hline
    \hline
    Model & Parameter &  Min & Max & N & Unit\\
    \hline
    \texttt{pow}& $\mathrm{Var}(\log(F_\mathrm{PL}))$    & 0.01 & 1 &10\\
    &$C_\Gamma$ &   0   &   2   &   10\\
    &Total   &   &   & 100\\
    \hline
    \texttt{bbdamp} &$f_\mathrm{BB}$    & 0 & 2 & 10\\
    &$kT$ &   0.01   &   1   &   10 & keV\\
    &Total   &   &   & 100\\
    \hline
    \texttt{refdamp}&$f_\mathrm{ref}$    & 0 & 5 & 10\\
    &$C_\mathrm{ref}$    & 0 & 1 & 10\\
    &$n$    & 15 & 19 & 5 & cm$^{-3}$\\
    &$\log(\xi)$    & 1 & 3 & 3 & erg~cm~s$^{-1}$\\
    &Total   &   &   & 1500\\
    \hline
    \texttt{ufo}&$C_\mathrm{UFO}$    & 0 & 1 & 10\\
    &$N_\mathrm{H}$    & 0.1 & 10 & 5 & $10^{23}$~cm$^{-2}$\\
    &$\log(\xi)$    & 3 & 5 & 3 & erg~cm~s$^{-1}$\\
    &Total   &   &   & 150\\
    \hline
    \hline
    \end{tabular}
    \label{tab:fvar_pow}
\end{table}

The parameters of the final \textsc{xspec} table models are shown in Table~\ref{tab:fvar_pow}, along with the number of steps in each parameter.

\section{Comparison with analytic calculation}
\label{sec:analytic}

 In this section, we consider a comparison between our Monte-Carlo derived models and the analytic solution for a variable power-law and constant reflection. We assume no correlations between parameters and constant $\Gamma$ and reflection flux, so that the analytical solution can be easily derived.

In this case, where the data can be described as an energy dependent constant component with flux $F_\mathrm{c}$ and a variable component with a normalization $N_\mathrm{v}$, normalization variance $\sigma_v^2$ and an energy dependent flux $F_\mathrm{v}$, the fractional excess variance be be described as:

\begin{equation}
    F_\mathrm{var}(E) = \frac{\sigma_v}{\bar{N}_\mathrm{v}}
    \left(1-\frac{F_\mathrm{c}(E)}{\bar{F}(E)}\right)
\end{equation}
where $F(E)$ is the total flux \citep{Vaughan04}. For this test, we use a density of $10^{17}$~cm$^{-3}$ and ionization of 100~erg~cm~s$^{-1}$, a reflection fraction of 1, and keep the other parameters are the same as assumed for the \textsc{fvar\_refdamp} model.

\begin{figure}
    \centering
    \includegraphics[width=\linewidth]{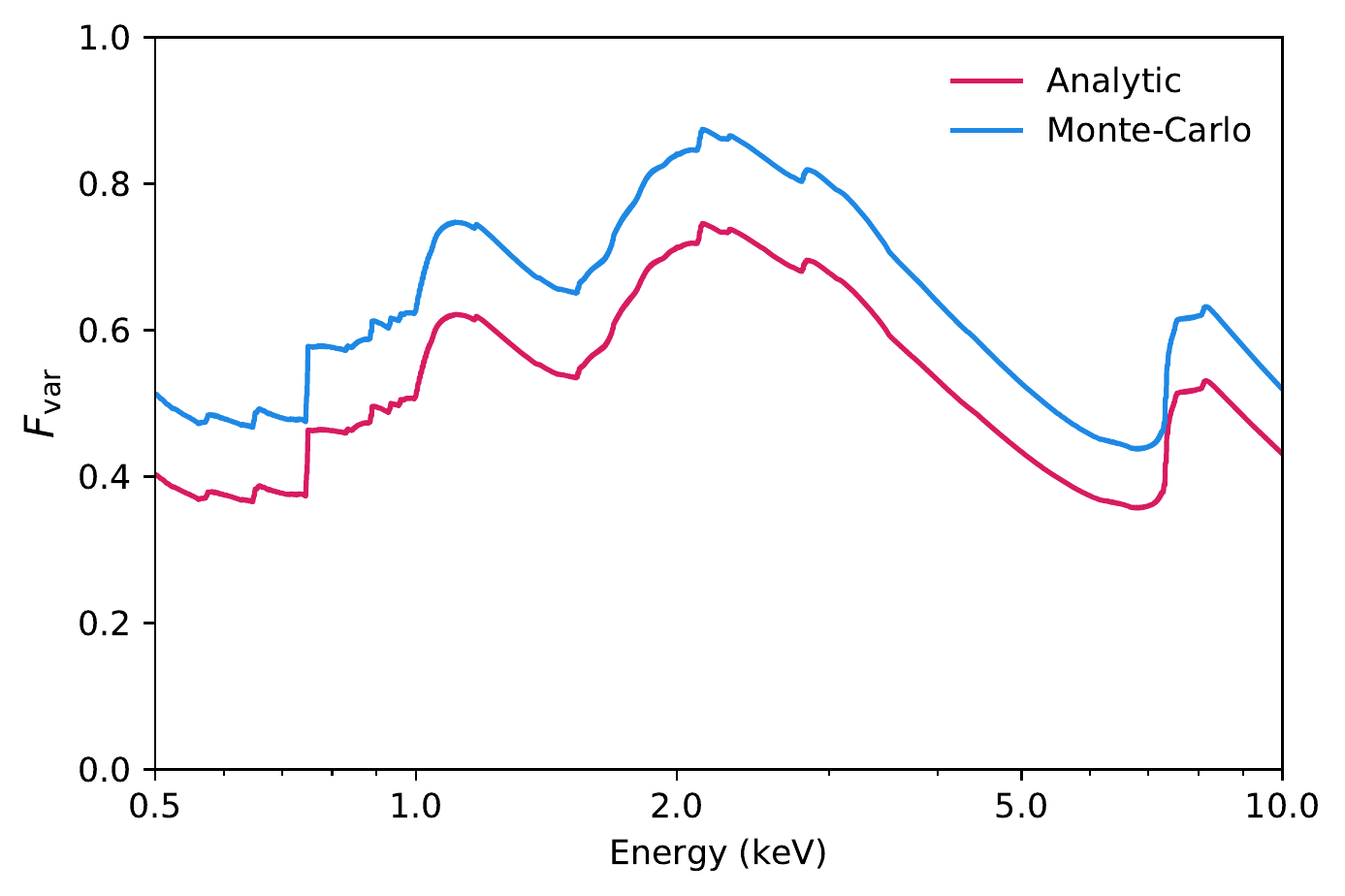}
    \caption{Variance spectra for a variable power-law and constant reflection, calculated using our models and using an analytic approach. The parameters are identical between the two, except that we use a slightly higher variance in the Monte-Carlo model so that the two lines can be distinguished.}
    \label{fig:analytic}
\end{figure}

In Fig.~\ref{fig:analytic} we show a comparison of the two models, which are an excellent match. We conclude that the Monte-Carlo approach can easily replicate the analytic solution, while also being much easier to expand to more complex models.

\section{The effect of instrumental response}
\label{sec:response}

Because we want our models to be applicable to X-ray data in general, we did not include a specific instrumental response when calculating them. However, while this means that the models can in principle be used with any spectra, it also makes them less accurate.

\begin{figure}
    \centering
    \includegraphics[width=\linewidth]{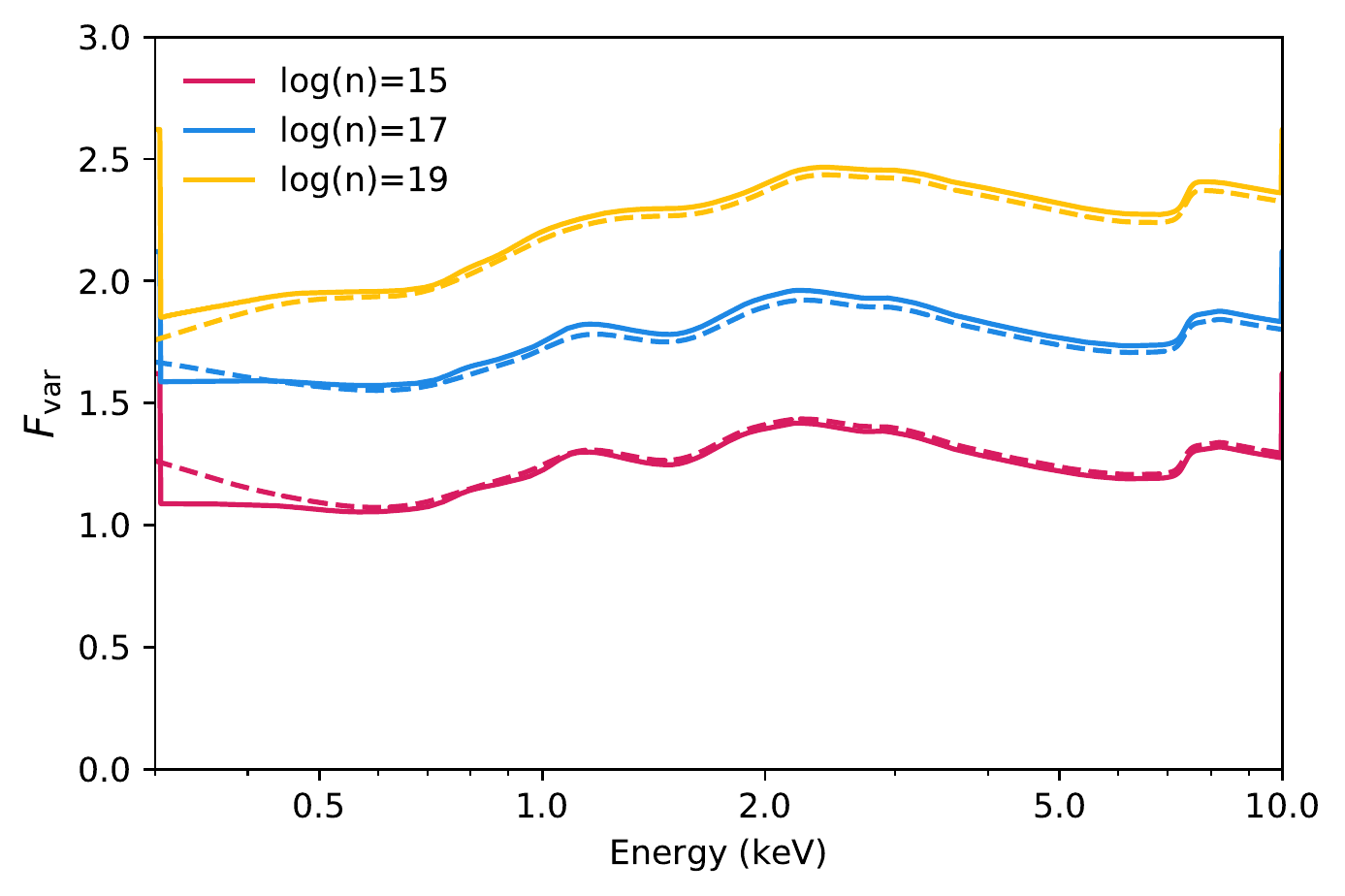}
    \caption{Comparison of Gaussian smoothed variance models with those calculated  using the EPIC-pn response. The model is a variable powerlaw with constant reflection, with three different values of the density parameter.}
    \label{fig:response}
\end{figure}

For this paper, we assume a Gaussian smoothing to mimic the spectral resolution of the EPIC-pn. To estimate the affect that this has on our calculated spectra, we recalculate the \textsc{fvar\_refdamp} model, this time with the input spectra convolved with the EPIC-pn response. In Fig.~\ref{fig:response} we show a comparison of this model with the Gaussian smoothed version, for three values of the density parameter.

The models are in excellent agreement, with the only difference being below 0.5~keV where they differ by $\sim10\%$. For our purposes, where we fit the soft excess with a semi-phenomenological model and are mainly interested in narrow features at higher energies, this level of accuracy is easily sufficient. However, for more sophisticated modelling of the soft X-ray band it may be necessary to use a more sophisticated treatment of the instrumental response.


\bsp	
\label{lastpage}
\end{document}